\documentclass[12pt]{article}
\usepackage{amsfonts}
\usepackage{graphicx}
\usepackage{amsmath}
\usepackage{hyperref}
\usepackage[cp866]{inputenc}
\begin{document}

\begin{center}
\textbf{Limiting Approach to Generalized Gamma Bessel
Model via Fractional Calculus and its Applications in Various Disciplines }

\bigskip

NICY SEBASTIAN

\vspace{3mm}Indian Statistical Institute, Chennai Centre, SETS Campus, MGR Knowledge City

CIT Campus, Taramani, Chennai, India-600 113.

e-mail: nicy@isichennai.res.in; nicyseb@yahoo.com
\bigskip
\end{center}

\begin{abstract}
The essentials of fractional calculus according to different approaches that can be useful
 for our applications in the theory of probability and stochastic processes are established. In addition to
 this, from this fractional integral one can list out
almost all the extended densities for the pathway parameter $q < 1$ and $q \rightarrow 1$. Here we bring out the idea of thicker or thinner tailed models associated with a gamma
type distribution as a limiting case of pathway operator.
Applications of this extended gamma model in Statistical Mechanics, input-output models, and solar spectral irradiance modeling  etc are established.
\end{abstract}
\vskip.3cm
\section{Introduction}
\vskip.3cm
In recent years considerable interest has been shown in the so-called fractional calculus, which allows one to consider integration and differentiation of any order, not necessarily integer.
Fractional calculus is a rapidly growing field both in
theory and in applications to real world problems. There is a
revived interest in fractional integrals and fractional derivatives
due to their recently found applications in reaction, diffusion,
reaction-diffusion problems, in solving certain partial differential
equations, in input-output models and related areas, see for example
\cite{Haubold and Mathai (2000)}, \cite{Henry  and Wearne (2002)}, \cite{KSS (2004)}, \cite{MLP (2001)},  \cite{Mathai  and Haubold (2007)},
\cite{SMH1(2006)}. There are several books in the
area, some of them are \cite{Kilbas  and Saigo (2004)}, \cite{Kiryakova (1994)}, \cite{Miller and Ross (1993)}, \cite{Podlubny (1999)} and \cite{SKM (1990)}.
The classical left and right-hand sided {{Riemann-Liouville
fractional integral operators}} of order $\alpha \in \mathbb{C},\
\Re(\alpha)>0,$ are defined as:
\begin{equation}\label{eqj:1}
{_0D}_x^{-\alpha}f=({I}_{0+}^{\alpha}f)(x)
  =\frac{1}{\Gamma(\alpha)}\int_o^x (x-t)^{\alpha-1}f(t)\rm{d}t,\ \
  x>0,\Re(\alpha)>0,
  \end{equation}
\begin{equation}\label{eqj:2}
 {_xD_{\infty}^{-\alpha}}f=({I}_{-}^{\alpha}f)(x)
=\frac{1}{\Gamma(\alpha)}\int_x^\infty (t-x)^{\alpha-1}f(t)\rm{d}t,\
\ x>0,\Re(\alpha)>0.
\end{equation}
\vskip.2cm
The traditional special functions are also related to the classical Fractional Calculus (FC), and later to the generalized fractional calculus, and shown to be representable as fractional
order integration or differentiation operators of some basic elementary functions. Such relations provided some alternative definitions for the special
functions by means of Poisson-type and Euler-type integral representations and
Rodrigues type differential formulas. Example of such an unified approach
on special functions, based on a generalized fractional calculus, can be seen in
\cite{Kiryakova (1994)}. The essentials of fractional calculus according to different approaches that can be useful for our applications in the theory of probability and stochastic processes are established with the help of pathway idea of \cite{Mathai (2005)}.

\vskip.2cm
The pathway idea was originally proposed by Mathai in the 1970's in connection with population models, and later rephrased and extended in \cite{Mathai (2005)} to cover scalar as well as matrix cases, and made suitable for modelling data from statistical and physical situations. The main idea behind the derivation of this model is the switching
 properties of going from one family of functions to another and yet another family of functions.
 It is shown that through a parameter
$q$, called the pathway parameter, one can connect generalized type-1 beta family
of densities, generalized type-2 beta family of densities, and generalized gamma
family of densities, in the scalar as well as in the matrix cases, also in the real and
complex domains. It is shown that when the model is applied to physical situations
then the current hot topics of Tsallis statistics and superstatistics in statistical mechanics
become special cases of the pathway model, and the model is capable of
capturing many stable situations as well as the unstable or chaotic neighborhoods
of the stable situations and transitional stages.  \cite{Mathai (2005)} deals
mainly with rectangular matrix-variate distributions and the scalar case is
a particular case there. For the real scalar case the pathway model is the
following:
\begin{equation}\label{eq:2.26}
h_1(x)=k_1x^{\gamma-1}[1-a(1-q)x^{\theta}]^{\frac{\eta}{1-q}},
 1-a(1-q)x^{\theta}>0, a, \theta, \gamma, \eta>0, q<1,
\end{equation}
where $k_1$
is the normalizing constant if a statistical density is needed. For $q < 1$ the model remains as a generalized
type-1 beta model in the real case.
Other cases available are the regular
type-1 beta density, Pareto density, power function, triangular and related
models. Observe that (\ref{eq:2.26}) is a model with the right tail cut off. When $q> 1$
we may write $1 - q = -(q - 1),~ q > 1$ so that $h_2(x)$ assumes the form,
\begin{equation}\label{eq:2.27}
h_2(x) = k_2x^{\gamma-1}
[1 + a(q - 1)x^\theta]^{-\frac{\eta}{q-1}}, x \geq 0, a, \theta, \gamma, \eta>0, q>1,  \end{equation}
which is a generalized type-2 beta model for real $x$ and $k_2
$ is the normalizing constant, if a statistical density is required.  Beck and Cohen's superstatistics
belong to this case (\ref{eq:2.27}) and dozens of published papers are available on the topic of superstatistics in astrophysics. For
$\gamma= 1,~ a = 1, \theta = 1$ we have Tsallis statistics for $q > 1 $ from (\ref{eq:2.27}). Other
standard distributions coming from this model are the regular type-2 beta,
the F-distribution, L$\acute{e}$vi models and related models. When $q\rightarrow 1,$ the forms
in (\ref{eq:2.26}) and (\ref{eq:2.27}) reduce to

\begin{equation}\label{eq:2.28}
h_3(x) = k_3x^{\gamma-1}{\rm e^{-b x^{\theta}}}
,~ x \geq0, b=a\eta>0, \gamma, \theta>0,
\end{equation}
where $k_3$
is the normalizing constant.
This includes generalized gamma, gamma, exponential, chisquare, Weibull,
Maxwell-Boltzmann, Rayleigh, and related models, for more details see  \cite{Honerkamp
(1994)} and \cite{Mathai(1993)}. If $x$ is replaced by $|x|$ in (\ref{eq:2.26}) then more families of distributions
are covered in (\ref{eq:2.26}).
Note that $q$ is the most important parameter here which enables one to move from one family of functions to another family. The other parameters are the usual parameters within each family of functions. %

\vskip.2cm The paper is organized as follows: In Section 2 the
connections of fractional integral operators to statistical
distribution theory and incomplete integrals are given. Section 3
covers limiting approach to generalized gamma model via pathway
operator. Application of extended generalized gamma
model in statistical mechanics is introduced in Section 4.
Generalized Laplacian density and stochastic process is introduced
in Section 5. In Section 6, we consider the application of
generalized gamma model in solar spectral irradiance modeling.
\vskip.3cm
\section{Statistical Interpretations
of Fractional Integrals}\vskip.3cm
A general pathway fractional integral operator is discussed in \cite{Nair (2009)}, which generalizes the classical Riemann-Liouville fractional integration
operator. The pathway fractional
integral operator has found applications in reaction-diffusion problems, non-extensive
statistical mechanics, non-linear waves, fractional differential equations, non-stable
neighborhoods of physical system etc.
By means of pathway model \cite{Mathai (2005)}, pathway
fractional integral operator (pathway operator) is defined
as follows:
Let $f(x)\in L(a, b), \eta\in C, \Re(\eta)>0, a>0$
and $q<1,$ then
\begin{equation}\label{eq:4.1}
(P_{0+}^{(\eta,q)}f)(x)=x^{\eta-1}\int_0^{\frac{x}{a(1-q)}}\left[1-\frac{a(1-q)t}{x}\right]^{\frac{\eta}{(1-q)}-1} f(t){\rm d}
t,\end{equation}
where $q$ is the pathway parameter and $f(t)$ is an arbitrary function. For more details,
see 
[36, 37].
In the pathway model, as $q\rightarrow1$, we can see that both $h_1(x)$ and $h_2(x)$ go to
$h_3(x)$ because
\begin{eqnarray*}
\lim_{q\rightarrow1_{-}}[1-a(1-q)x^{\delta}]^{\frac{\eta}{1-q}}
&=
&\lim_{q\rightarrow1_{+}}[1+a(q-1)x^{\delta}]^{-\frac{\eta}{q-1}} \\
   &=& {\rm e}^{-a\eta x^{\delta}}.
\end{eqnarray*}

When $q\rightarrow  1_{-}, [1-\frac{a(1-q)t}
{x} ]^{\frac{\eta}{1-q}}
\rightarrow {\rm e}^{-\frac{a\eta}{x}t}$. Thus the operator will become
$$P_{0+}^{\eta,1}=x^{\eta-1}\int_0^\infty{\rm e}^{\frac{-a\eta}{x}t} f(t){\rm d} t=x^{\eta-1}L_f(\frac{a\eta}{x}), $$
the Laplace transform of $f$ with parameter $\frac{a\eta}{x}$. When $q = 0, a = 1$ in (\ref{eq:4.1}) the integral
will become,
$$\int_0^x(x-t)^{\eta-1}f(t){\rm d} t=\Gamma(\eta)I_{0+}^\eta,$$
where $I_{0+}$ is the left-sided Riemann-Liouville fractional integral operator.

\vskip.2cm
 Fractional integrals in the matrix-variate
cases and their connection to statistical distributions are
 pointed out in \cite{Mathai (2009)} and \cite{Nair (2011)}.
Let $x>0$ and $y>0$ be statistically independently distributed
positive real scalar random variables. Let the densities of $x$ and
$y$ be $f_1(x)$ and $f_2(y)$ respectively. Then the joint density of
$x$ and $y$ is $f(x,y)=f_1(x)f_2(y)$. Let $u=x+y,t=y$. Then the
density of $u$, denoted by $g_1(u)$, is given by
\begin{equation}\label{eq:3}g_1(u)=\int_{t=0}^uf_1(u-t)f_2(t){\rm d}t
.\end{equation}Here (\ref{eq:3}) is in the same format of Riemann-Liouville
left-sided fractional integral for $f_1(x)=c_1~x^{\alpha-1}$ and
$f_2(y)=c_2f(y)$ where $c_1$ and $c_2$ are normalizing constants to
create densities. Thus a constant multiple of the left-sided
Riemann-Liouville fractional integral can be interpreted as the
density $g_1(u)$ of a sum of two independently distributed real
positive scalar random variables.

\vskip.2cm
 Now let us look at $u=x-y$ with the additional assumption
that $u=x-y>0$. Then the density of $u$, denoted by $g_2(u)$,will
have the format
\begin{equation}\label{eq:4}g_2(u)=\int_{t=u}^{\infty}f_1(t)f_2(t-u){\rm d}t.
\end{equation}By taking $f_2(y)=c_2y^{\alpha-1}$ and
$f_1(x)=c_1f(x)$, where $c_1$ and $c_2$ are some normalizing
constants, (\ref{eq:4}) agrees with the density of a structure $u=x-y$ with
$x-y>0, x>0,y>0$. Thus the right-sided Riemann-Liouville fractional
integral can be given the interpretation as the density of
$u=x-y>0,x>0,y>0$ where $x$ and $y$ are statistically independently
distributed real scalar random variables. \vskip.2cm
\vskip.2cm
Let us look into some examples from \cite{Mathai (2009)} and \cite{Sebastian (2009a)}. A real
positive scalar random variable $x$ is said to have a gamma density
if its density function is of the form

$$f(x)=\frac{ m^{\alpha}}{\Gamma(\alpha)}
x^{\alpha-1}{\rm
e}^{{-{m}} x},0\leq x<\infty,\alpha>0,{m}>0
$$and $f(x)=0$ elsewhere. Here $f(x)\ge 0$ for all $x$ and
$\int_{-\infty}^{\infty}f(x){\rm d}x=1$ making $f(x)$ a statistical
density. In this case
$$1=\frac{{{m}}^{\alpha}}{\Gamma(\alpha)}\int_0^{\infty}
t^{\alpha-1}{\rm
e}^{-{m} t}{\rm d}t.
$$Let us take a fraction of this integral such as ${\rm e}^{-{{a} x}}$
times this total integral $1$. That is,
\begin{eqnarray*}{\rm e}^{-{m} x}(1)&=&{\rm
e}^{-{m} x}\frac{{m}^{\alpha}}{\Gamma(\alpha)}\int_0^{\infty}
t^{\alpha-1}{\rm
e}^{- {m} t}
{\rm d}t\nonumber\\
&=&{{{m}^{\alpha}}\over{\Gamma(\alpha)}}\int_0^{\infty}t^{\alpha-1}{\rm
e}^{-{m}(t+x)}{\rm d}t, u=t+x\nonumber\\
&=&{{{m}^{\alpha}}\over{\Gamma(\alpha)}}\int_{u=x}^{\infty}(u-x)^{\alpha-1}{\rm
e}^{-u}{\rm d}u\nonumber\\
 &=&{m}^{\alpha}({I}_{-}^{\alpha}f)(x)\hbox{  with
}f(u)={\rm e}^{-{m} u}. \end{eqnarray*}Thus the constant multiple of right-sided
Riemann-Liouville fractional integral when $f(u)={\rm e}^{-{m} u}$ can
be interpreted as a fraction of the total integral coming from a
gamma density. Let us examine a fraction of the type-1 beta density.
A real scalar random variable $u$ is said to have a type-1 beta
density if the density function is given by
$$f(u)={{u^{\alpha-1}(1-u)^{\beta-1}}\over{B(\alpha,\beta)}},
 0\leq u<1,\alpha>0,\beta>0
$$and zero elsewhere, where
$B(\alpha,\beta)=\Gamma(\alpha)\Gamma(\beta)/\Gamma(\alpha+\beta)$.
The total probability in this case is given by
$$1=\int_0^1{{u^{\alpha-1}(1-u)^{\beta-1}}\over{B(\alpha,\beta)}}{\rm
d}u.
$$Let us consider a fraction of this total probability and consider
$b^{\alpha+\beta-1}(1)$. That is,
\begin{eqnarray*}b^{\alpha+\beta-1}&=& b^{\alpha+\beta-1}
\int_0^1{{u^{\alpha-1}(1-u)^{\beta-1}}\over{B(\alpha,\beta)}}{\rm
d}u\nonumber\\
&=&\int_0^b{{(b-t)^{\alpha-1}t^{\beta-1}}\over{B(\alpha,\beta)}}{\rm
d}t\nonumber\\
&=&{{\Gamma(\alpha+\beta)}\over{\Gamma(\beta)}}({I_{0+}^\alpha}f)(x),\
\
f(t)=t^{\beta-1}. \end{eqnarray*}
Thus the left-sided
Riemann-Liouville fractional integral when $f(t)={t}^{\beta-1}$ can
be interpreted as {a fraction of the
total integral coming from a beta density.

\vskip.2cm
Similarly a constant multiple of the left-sided pathway fractional integral can be interpreted as the
density of a sum of two independently distributed real positive scalar random variables, see \cite{Nair (2011)}.
Let $x > 0$ and $y > 0$ be statistically independently distributed
positive real scalar random variables with densities $f_1(x)$ and $f_2(y)$, respectively. Let $u = x +
a(1-q)y, t = y$. Then the density of $u$ is given by
\begin{equation}\label{eq:4.4}g_3(u)=\int_{t=0}^{\frac{u}{a(1-q)}}f_1(u-a(1-q))f_2(t){\rm d}t.
\end{equation}
This is in the same format of left-sided pathway fractional integral for $f_1(x)=c_1 (\frac{x}{u})^{\frac{\eta}{(1-q)}-1}$ and $f_2(y)=c_2 u^{\eta-1} f(y)$.
That is
\begin{eqnarray*}g_3(u)
&=&c_1c_2x^{\eta-1}\int_0^{\frac{x}{a(1-q)}}\left[1-\frac{a(1-q)t}{x}\right]^{\frac{\eta}{(1-q)}-1} f(t){\rm d}
t\nonumber\\
&=&c_1c_2P_{0+}^{(\eta,q)}. \end{eqnarray*}
Likewise
statistical interpretations can be given for other fractional integrals also. If we replace $f(t)$ by a non-negative integrable function, one can
obtain a statistical density through this operator. In addition to this, from this fractional integral one can list out
almost all the extended densities for the pathway parameter $q < 1$ and $q \rightarrow 1$, for more details see \cite{Nair (2011)}.

\vskip.3cm
\section{Limiting Approach to Generalized Gamma
Bessel Model via Pathway Operator}\vskip.3cm
Here we bring out the idea of thicker or thinner tailed models associated with a gamma
type distribution as a limiting case of pathway operator.
Let the integrand of (\ref{eq:4.1}) be denoted by $I_{(\eta,q)}$.
\begin{equation}\label{eq:4.2}
I_{(\eta,q)}=\left[1-\frac{a(1-q)t}{x}\right]^{\frac{\eta}{(1-q)}-1} f(t),~ \eta>0.\end{equation}
If we consider any real-valued positive integrable scalar function of $t$ instead of any
arbitrary real-valued scalar function of $t$, one can bring out a statistical density from
the pathway fractional integral operator. Thus one can say that
$$f_{q}(t)=CI_{(\eta,q)}(t) $$
is a statistical density. Hence (\ref{eq:4.1}) generalizes all the left-sided standard fractional
integrals and almost all the extended densities for $q < 1$ and $q \rightarrow 1$. In (\ref{eq:4.1}), when
$q\rightarrow 1$, the integrand $I_{(\eta,q)}$ will become
$$I_{(\eta,1)}={\rm e}^{-\frac{a\eta}{x}t} f(t).$$
In particular if we take $f(t) = 1$ and $\frac{a\eta}{x} = b > 0$, then one has obtained the Gaussian
or normal density. For $q\rightarrow1$, and $f(t)$ is replaced by $t^{\beta}$, we have the gamma density.
Similarly standard type-1 beta density, Pathway model for $q< 1$, chisquare density,
exponential density and many more can be obtained as a special case of pathway
integral operator. From (\ref{eq:4.2}) one can obtain the generalized gamma Bessel density as a limiting case. When $q\rightarrow 1_{-}$ and replace $f(t)$ by $t^{\beta-1 }{_0F_1}(~; \beta;\delta t)$,
then $g(t)$ will be
\begin{equation}\label{eq:4.3}
 g(t)=\left\{\begin{array}{ll}
C t^{\beta-1}{\rm e}^{-b t}{_0F_1}(~;\beta;\delta t);
 & t\geq0,~\beta, b>0\\
 0;&\text{otherwise.}
 \end{array} \right .
\end{equation}
\begin{figure}
\begin{center}
~~~~~ \resizebox{7cm}{5cm}{\includegraphics{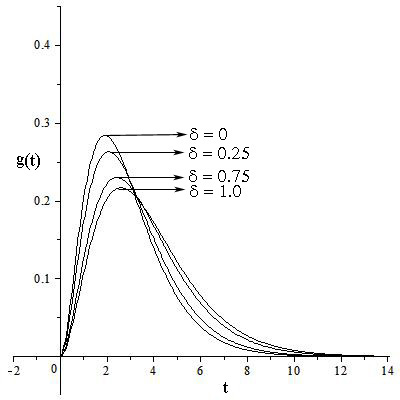}}
~~~~ \resizebox{7cm}{5cm}{\includegraphics{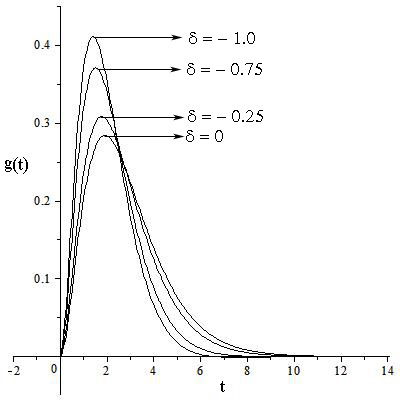}}\\
\caption {\small (a)
gamma Bessel model for  $\delta>0,$
\hskip1.5cm (b)gamma Bessel model for
$\delta<0$\label{ppplot1}}
\end{center}
\end{figure}
Some of the special cases of (\ref{eq:4.3}) are given in Table \ref{table21}.
\begin{table}
{\caption{\small Special cases of generalized
gamma model associated with Bessel function \label{table21}} }
\begin{center}
{\scriptsize{
\begin{tabular}{|l|l|}
 \hline
 &\\
 $\delta=0$&\text{Two
 parameter gamma density}\\
  \hline
 &\\
$\delta=0,a=1$&\text{One
 parameter gamma density}\\
  \hline
 &\\
$\delta=0,\beta=1$&\text{Exponential
 density}\\
  \hline
 &\\
$\delta=0,a=\frac{1}{2},\beta=\frac{n}{2}, n=1,2,\cdots$&\text{Chi-square density}\\
 \hline
 & \\
 $\delta=\lambda,a=\frac{1}{2},\beta=\frac{n}{2}, n=1,2,\cdots$
& \text{Noncentral chi-square density}\\
\hline
\end{tabular}}}
\end{center}
\end{table}
For fixed values of $\beta$ and $b$, we can look at the graphs for
$\delta>0$ as well as for $\delta<0$. These graphs give a suitable
interpretation, when tail areas are considered.
In Figure \ref{ppplot1}(a), note
that $\delta=0$ is the case of a gamma density. Thus when $\delta$
increases from $\delta=0$ the right tail of the density becomes
thicker and thicker. Thus, when fitting a gamma type model to given
data and if it is found that a model with a thicker tail is needed
then one can select a member from this family for appropriate
$\delta>0$. In Figure \ref{ppplot1}(b), observe that $\delta=0$ is the case of gamma density. When $\delta$ decreases from  $\delta=0$ the right tail gets thinner
and thinner. Thus if we are looking for a gamma type density but
with a thinner tail then one from this family may be appropriate for
$\delta<0$. For more details of the model in (\ref{eq:4.3}), see \cite{Sebastian (2009)} and \cite{Sebastian (2011)}. When $ q\rightarrow1_{-}, \eta=1$ and replace $f(t)$ by $t^{\beta-1 }{_0F_1}(~; \beta;\delta t)$ in pathway fractional integral operator then we are essentially dealing with
distribution functions under a gamma Bessel type
model in a practical statistical problem. Which provides a connection between
statistical distribution theory and fractional calculus so that one can make use of
the rich results in statistical distribution theory for further development of fractional
calculus and vice versa.
\vskip.2cm
We can look at the model in another way also. Consider the total integral as
$$1 = C \int_0^{\infty}t^{\beta-1}{\rm e}^{-b t}{_0F_1}(~;\beta;\delta t){\rm d}t $$
which can be treated as the Laplace transform of the function $t^{\beta-1 {_0F_1}(~; \beta;\delta t)},$ and
hence $C=\frac{b^{\beta}}{\Gamma(\beta){\rm e}^{\frac{\delta}{b}}}$ where $C$, the normalizing constant of (\ref{eq:4.3}), is nothing but the
Laplace transform of the given function. It is shown to be very relevant in fractional
reaction-diffusion problems in physics. Similarly for $b = 0$, it will become the Mellin transform of the function
${_0F_1}(~;\beta;\delta t)$.
\vskip.2cm
\begin{figure}
\begin{center}
~~~~~ \resizebox{7cm}{5cm}{\includegraphics{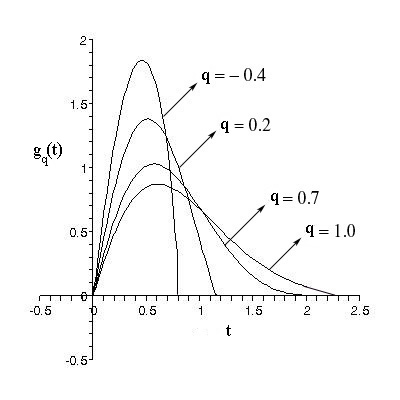}}
~~~~ \resizebox{7cm}{5cm}{\includegraphics{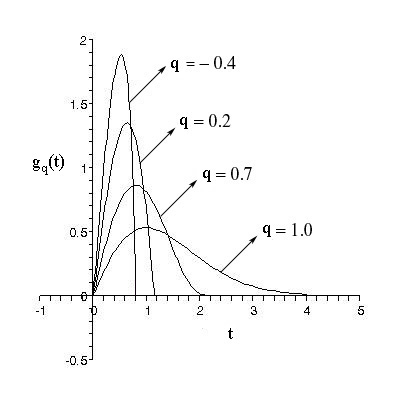}}\\
\caption{\small (a) The $q$-gamma Bessel model for $q<1, \delta=-0.50$ \hskip.5cm (b) The $q$-gamma Bessel model for $q<1, \delta=0.5$.  \label{ppplot2}}
\end{center}
 \end{figure}
 \begin{figure}
\begin{center}
\resizebox{12cm}{5cm} {\includegraphics{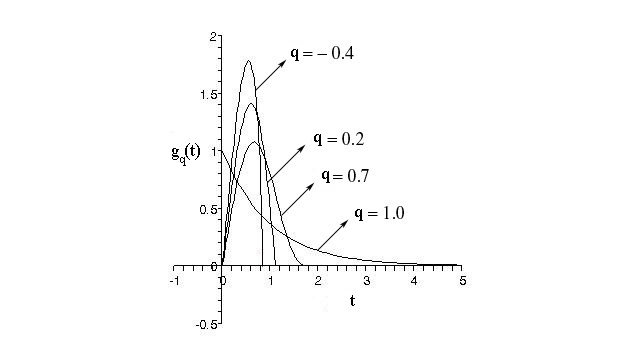}}\\
\caption{\small The $q$-gamma Bessel model for $q<1, \delta=0$.\label{ppplot3}}
\end{center}
\end{figure}
The $q$-analogue of generalized gamma Bessel density can also be deduced from the pathway fractional integral operator, by putting $x=1, \eta=1$ and replace $f(t)$ by $t^{\beta-1 }{_0F_1}(~; \beta;\delta t)$,
then $g_{q}(t)$ will be
\begin{equation}\label{eq:4.3s}
g_{q}(t)=\left\{\begin{array}{ll}
K t^{\beta-1}[1-b(1-q)t]^{\frac{1}{1-q}}{_0F_1}(~;\beta;\delta t);
 &q<1,1-b(1-q)t>0, t>0,\beta,b>0\\
 0;&\text{otherwise,}
 \end{array} \right .
\end{equation}
where $K$ is the normalizing constant.
For fixed values of $b$ and $\beta$, we can look at the graphs for
$\delta=-0.5, q<1$, $\delta=0.5, q<1$ as well as for $\delta=0, q<1$.
From the Figures \ref{ppplot2} and \ref{ppplot3}, we can see that when $q$ moves from -1
to 1, the curve becomes thicker tailed and less peaked.
It is also observed that
when $\delta>0$
the right tail of the density becomes
thicker and thicker. Similarly when $\delta<0$ the right tail gets thinner
and thinner.
Observe that for $q>1,$ writing
$1-q=-(q-1)$ in equation (\ref{eq:4.3}) produces extended type-2
beta form which is given by
\begin{equation}\label{eq:4.3s}
f_{q}(t)=\left\{\begin{array}{ll}
P t^{\beta-1}[1+b(q-1)t]^{-\frac{1}{q-1}}{_0F_1}(~;\beta;\delta t);
 &q>1, t>0,~\beta, b>0\\
 0;&\text{otherwise,}
 \end{array} \right .
\end{equation}
where $P$ is the normalizing constant. From Figure \ref{ppplot41},  we can see that when $q$ moves from 1
to $\infty$, the curve becomes less peaked.
In this case also it is observed that
when $\delta>0$
the right tail of the density becomes
thicker and thicker and when $\delta<0$ the right tail gets thinner
and thinner.
\begin{figure}
\begin{center}
\resizebox{7cm}{5cm}{\includegraphics{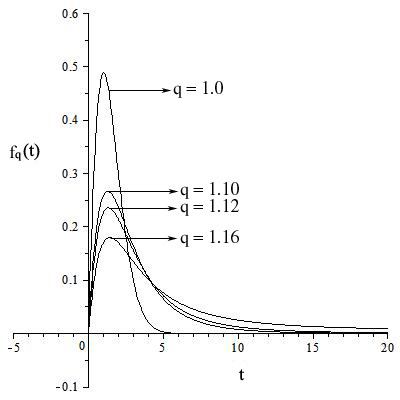}}
\resizebox{7cm}{5cm}{\includegraphics{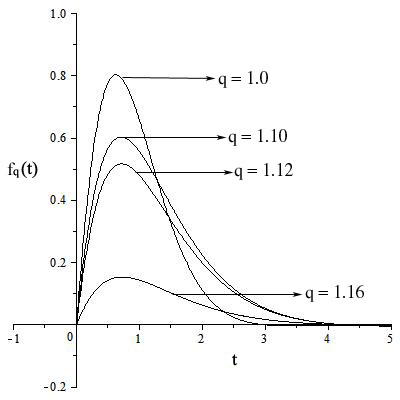}}\\
\caption {\small (a)
$q$-gamma bessel model for $\delta=0.5,q>1$
\hskip1cm (b) $q$-gamma bessel model for $\delta=-0.5,q>1$\label{ppplot41}}
\end{center}
\end{figure}
\vskip.3cm
 Densities exhibiting thicker or thinner tail occur frequently in many different
areas of science. For practical
purposes of analyzing data from physical experiments and in building up models in statistics,
we frequently select a member from a parametric family of distributions. But it is often found
that the model requires a distribution with a thicker or thinner tail than the ones available
from the parametric family.
\vskip.3cm
\section{Applications in Statistical Mechanics}\vskip.3cm
Nonequilibrium complex systems often exhibit dynamics that can be decomposed
into several dynamics on different time scales. As a simple example, consider a
Brownian  motion of a particle moving through a changing fluid environment, characterized by
temperature variations on a large scale. In this case, two dynamics are relevant: one is a
fast dynamics describing the local motion of the Brownian particle and the other one is
a slow one due to the large global variations of the environment with spatio-temporal
inhomogeneities. These effects produce a superposition of two different statistics, which
is referred to as superstatistics. The concept of superstatistics has been introduced by \cite{Beck and Cohen (2003)}  and \cite{Beck (2006)}
after some preliminary considerations in \cite{Beck (2001)} and
\cite{Beck (2002)}. The stationary distributions of
superstatistical systems typically exhibit a non-Gaussian behavior with fat tails, which
can decay, for example, as a power law, a stretched-exponential law, or in an even more
complicated way \cite{Touchette (2004)}. Essential for this approach is the existence of an intensive variable say $\beta$, which fluctuates on a large spatio-temporal scale.
\vskip.2cm
For the above-mentioned example of a superstatistical Brownian particle, $\beta$ is the
fluctuating inverse temperature of the environment. In general, however, $\beta$ may also be
an effective friction constant, a changing mass parameter, a variable noise strength, the
fluctuating energy dissipation in turbulent flows, a fluctuating volatility in finance, an
environmental parameter for biological systems, a local variance parameter extracted
from a signal, and so on. Superstatistics offers a very general framework for treating nonequilibrium stationary
states of such complex systems. After the original work in \cite{Beck and Cohen (2003)}, a lot of efforts have
been made for further theoretical elaboration, see \cite{Beck (2004)}  and 
 \cite{Chavanis (2006)}. At the same time, it has also been
applied successfully to a variety of systems and phenomena, including hydrodynamic
turbulence, 
 pattern formation,
 cosmic rays,
mathematical finance,
 random matrices, 
  and hydro-climatic fluctuations. 

\vskip.2cm
 From a statistical point of view, the procedure is equivalent
 to starting with a conditional distribution of a gamma type for every given value
  of a parameter $a$. Then $a$ is assumed to have a prior known density of the gamma type.
  Then the unconditional density is obtained by integrating out over the density of $a$.
Let us consider the conditional density of the form
\begin{equation}\label{eq:2.14}
f_{x|a}(x|a)=k_1 x^{\gamma-1}{\rm
e}^{-ax^{\rho}}{_0F_1}(~;\frac{\gamma} {\rho};\delta x^\rho);~0\leq
x<\infty,\rho,a,\gamma>0
\end{equation}
and $f(x)=0$ elsewhere, where $k_1$ is the normalizing constant. When $\delta=0$
the equation (\ref{eq:2.14}) reduces to generalized gamma density.
Note that this is the generalization of some standard statistical
densities such as gamma, Weibull, exponential, Maxwell-Boltzmann,
Rayleigh and many more. When we put $\rho=1$ in equation
(\ref{eq:2.14}), it reduces to (\ref{eq:4.3}). When $\delta=0,\rho=2$
(\ref{eq:2.14}) reduces to folded standard normal density.
\vskip.2cm
Suppose that $a$ has a gamma density given by
\begin{equation}\label{eq:2.15}
f_a(a)=\frac{\lambda^\eta a^{\eta-1}{\rm e}^{-\lambda a}}
{\Gamma(\eta)};~0\leq a<\infty,\eta,\lambda>0,
\end{equation}
and $f_a(a)=0$ elsewhere. In a physical problem, the residual rate of change may have small
probabilities of it being too large or too small and the maximum
probability may be for a medium range of values for the residual
rate of change $a$. This is a reasonable assumption. Then the
unconditional density of $x$ is given by

\begin{equation}\label{eq:2.16}
f_x(x)=\int_a f_{x|a}(x|a)f_a(a){\rm d} a =\frac{\rho\lambda^\eta
x^{\gamma-1}}{\Gamma(\frac{\gamma}{\rho})\Gamma(\eta)}{_0F_1}(~;\frac{\gamma}
{\rho};\delta x^\rho)~ I_{11}
\end{equation}
where \begin{equation}\label{eq:2.17}I_{11}=\int_0^\infty
a^{\frac{\gamma}{\rho}+\eta-1}{\rm e}^
{-a(\lambda+x^\rho)-\frac{\delta}{a}}{\rm d} a.\end{equation} Note
that one form of the inverse Gaussian probability density function
is given by
$$h_1(x)=c x^{-\frac{3}{2}}\rm e^{-\frac{\xi}{2}[\frac{x}{\nu^2}+\frac{1}{x}]},\ \ \nu\neq0,\xi>0,x\geq0,$$
where $c$ is the normalizing constant. Put
$\frac{\gamma}{\rho}+\eta-1=-\frac{3}{2},
\lambda+x^\rho=\frac{\xi}{2\nu^2},\delta=\frac{\xi}{2}$ in $I_{11}$,
we can see that the inverse Gaussian density is the integrand in
$I_{11}$. Hence $I_{11}$ can be used to evaluate the moments of
inverse Gaussian density. Also $I_{11}$ is the special case of
reaction rate probability integral in nuclear reaction rate theory,
Kr\"{a}tzel integrals in applied analysis etc (see \cite{Joseph (2009)}, \cite{Kr(1979)}, \cite{Mathai and Haubold (1988)}, \cite{Mathai (2007)}).
For the evaluation of this integral and for more details see \cite{Sebastian (2009)} and \cite{Sebastian (2011)}.
Hence we have the unconditional density
\begin{equation}\label{eq:2.22}
f_x (x)=\frac{ \rho
\lambda^\eta}{\Gamma(\frac{\gamma}{\rho})\Gamma(\eta)}\frac{x^
{\gamma-1}}{(\lambda+x^{\rho})^{\frac{\gamma}{\rho}+\eta}}{_0F_1}(~;\frac{\gamma}
{\rho};\delta x^\rho)G_{0, 2}^{2, 0}\bigl[\delta(\lambda+x^\rho)
\big|_{0, \frac{\gamma}{\rho}+\eta }\bigr],
\end{equation}
where $G-$ function is defined as the following Mellin-Barnes
integral:

\begin{equation*}
G_{p, q}^{m, n}\bigl[z \big|_{b_1, \ldots,
b_q}^{a_1,\ldots,a_p}\bigr]=\frac{1}{2\pi
i}\int_{\mathcal{L}}\Phi(s)z^{-s}{\rm d} s,
\end{equation*}
where $$\Phi(s)=\frac{\bigl\{\prod _{j=1}^m \Gamma (b_j+
s)\bigr\}~\bigl\{\prod _{j=1}^n\Gamma (1-a_j- s)\bigr\}}
{\bigl\{\prod _{j=m+1}^q\Gamma (1-b_j- s)\bigr\} \bigl\{ \prod
_{j=n+1}^p \Gamma (a_j+ s)\bigr\}}$$ with $a_j, j=1,\ldots,p$ and
$b_j, j=1,\ldots,q$ are complex numbers, $\mathcal{L}$ is a contour
separating the poles of $\Gamma (b_j+ s), j=1,\ldots,m$ from those
of $\Gamma (1-a_j- s), j=1,\ldots,n.$ Convergence conditions,
properties and applications of $G-$ function in various disciplines
are available in the literature. For example, see \cite{Kilbas and Saigo (2004)}.
Equation (\ref{eq:2.22}) is a superstatistics, in
 the sense of superimposing another distribution or the distribution of $x$
with superimposed distribution of the parameter $a.$ In a physical
problem the parameter may be something like temperature having its
own distribution. Several physical interpretations of
superstatistics are available
 from the papers of Beck and others.
\vskip.2cm
We can easily obtain the series representation
of the unconditional density (\ref{eq:2.22}) and is given by
\begin{eqnarray}\label{eq:2.25}
f_x(x)&=&\frac{ \rho
\lambda^\eta}{\Gamma(\frac{\gamma}{\rho})\Gamma(\eta)}\frac{x^
{\gamma-1}}{(\lambda+x^{\rho})^{\frac{\gamma}{\rho}+\eta}}{_0F_1}(~;\frac{\gamma}
{\rho};\delta x^\rho)\sum_{k=0}^\infty\frac{\Gamma
(\frac{\gamma}{\rho}+\eta-k)(-1)^k [\delta(\lambda+x^\rho)]^k}{k!}\nonumber\\
&=&\frac{ \rho\Gamma(\frac{\gamma}{\rho}+\eta)
\lambda^\eta}{\Gamma(\frac{\gamma}{\rho})\Gamma(\eta)}\frac{x^
{\gamma-1}}
{(\lambda+x^{\rho})^{\frac{\gamma}{\rho}+\eta}}{_0F_1}(~;\frac{\gamma}
{\rho};\delta x^\rho){_0F_1}(~;1-\frac{\gamma}
{\rho}-\eta;\delta(\lambda+x^\rho))\nonumber\\
&&\lambda,\rho,\eta,\delta>0,\frac{\gamma}{\rho}>0,
1-\frac{\gamma}{\rho}-\eta\neq-\nu,\nu=0,1,\cdots, x\geq0.
\end{eqnarray}
This series representation provides an extension of the Beck and Cohen statistic. Thus, (\ref{eq:2.25}) gives a
suitable interpretation, when tail areas are shifted. This model has wide potential applications in  physical sciences especially in statistical mechanics, see \cite{Sebastian (2009)} and \cite{Sebastian (2011)}.
\vskip.3cm
\section{Applications in Growth-Decay Mechanism}\vskip.3cm
If $x$ is replaced by $|x|$ in (\ref{eq:2.26}) and when $q\rightarrow 1$, real scalar case of the pathway model takes the form,
\begin{equation}\label{eq:tt}
h_4(x)=c_4|x|^{\gamma-1}{\rm e}^{-a|x|^{\theta}}, -\infty <x<\infty, a>0.
\end{equation}
The density in (\ref{eq:tt}) for $\gamma=1, \theta=1$ is the simple Laplace
density. For $\gamma=1$ we have the symmetric Laplace density. A
general Laplace density is associated with the concept of
Laplacianness of quadratic and bilinear forms. For the concept of
Laplacianness of bilinear forms, corresponding to the chisquaredness
of quadratic forms, and for other details see \cite{Mathai(1993)} and
\cite{MPH (1995)}. Laplace density is also
connected to input-output type models. Such models can describe many
of the phenomena in nature. When two particles react with each other
and energy is produced, part of it may be consumed or converted or
lost and what is usually measured is the residual effect. The water
storage in a dam at a given instant is the residual effect of the water flowed into
the dam minus the amount taken out of the dam. Grain storage in a
sylo is the input minus the grain taken out. Hence it is of great importance in modeling this residual effect and lot of studies are there on this concept. There are several input-output type situations in economics, social sciences, industrial production, commercial activities, cosmological studies, and so on.
 It is shown in \cite{Mathai(1993a)} that when we have independently distributed gamma type input and gamma type output the
residual part $z=x-y, x=$ input variable, $y= $ output variable then
the special cases of the density of $z$ is a Laplace density. In
this case one can also obtain the asymmetric Laplace and generalized
Laplace densities, which are currently used very frequently in
stochastic processes, as special cases of the input-output model.
\vskip.2cm
Generalized gamma Bessel model in (\ref{eq:4.3}) has the moment generating function
$$
M_x(t) =\frac{{{b}}^{\beta_1}}{{\rm e}^{\frac{\delta_1}{b}}} \frac{{\rm
e}^{\frac{\delta_1}{b-t}}}{{(a_1-t)}^{\beta_1}}, b-t>0, \beta_1 >0.
$$
Let $x$ and $y$ be two independently distributed generalized gamma Bessel models having parameters
$(\alpha_1,\beta_1,\delta_1)$ and $(\alpha_2,\beta_2,\delta_2)$ respectively, $\alpha_i>0, \beta_i>0,\delta_i, i=1,2$. Let $z=x-y$. Due to the independence of $x$ and $y$ the moment generating function of $u$ is given by
$$M_z(t) =\frac{{{\alpha_1}}^{\beta_1}}{{\rm e}^{\frac{\delta_1}{\alpha_1}}} \frac{{\rm
e}^{\frac{\delta_1}{\alpha_1-t}}}{{(\alpha_1-t)}^{\beta_1}}\frac{{{\alpha_2}}^{\beta_2}}
{{\rm e}^{\frac{\delta_2}{\alpha_2}}} \frac{{\rm
e}^{\frac{\delta_2}{\alpha_2+t}}}{{(\alpha_2+t)}^{\beta_2}}, ~\alpha_1-t>0, \alpha_2+t >0.
$$
When $\alpha_1=\alpha_2=\alpha,~ \beta_1=\beta_2=\beta,~ \delta_1=\delta_2=\delta=0,$ then the above equation reduces to that of the generalized Laplacian model of Mathai.

\vskip.3cm
\section{Applications in Solar Spectral Irradiance Modeling}\vskip.3cm

Any object with a temperature above absolute zero emits radiation.
The Sun, our singular source of renewable energy, sits at the center of the solar system and emits energy as electromagnetic radiation at an extremely large and relatively constant rate, 24 hours per day, 365 days of the year.
With an effective temperature of approximately 6000 K, the sun emits radiation over a wide range of wavelengths, commonly labeled from high energy shorter wavelengths to lower energy longer wavelengths as gamma ray, x-ray, ultraviolet, visible, infrared and radio waves. These are called spectral regions.
The rate at which solar energy reaches a unit area at the earth is called the ``solar irradiance" or ``insolation".
Solar irradiance is an instantaneous measure of rate and can vary over time.
The units of measure for solar radiation are joules per square meter ($J/m^2$) but often watt-hours per square meter $(Wh/m^2)$ are used. As will be described above, solar radiation is simply the integration or summation of solar irradiance over a time period. For more details see \cite{Gueymard (2004)} and \cite{Stoffel et al. (2010)}. Good quality, reliable solar radiation data is becoming increasingly important in the field of renewable energy, with regard to both photovoltaic and thermal systems. It helps well-founded decision making on activities such as research and development, production quality control, determination of optimum locations, monitoring the efficiency of installed systems and predicting the system output under various sky conditions. Especially with larger solar power plants, errors of a very few percent can significantly impact upon the return on investment.
 Scientists studying climate change are interested in understanding the effects of variations in the total and spectral solar irradiance on Earth and its climate.

\begin{figure}[h!]
\begin{center}
\resizebox{7cm}{5cm}{\includegraphics{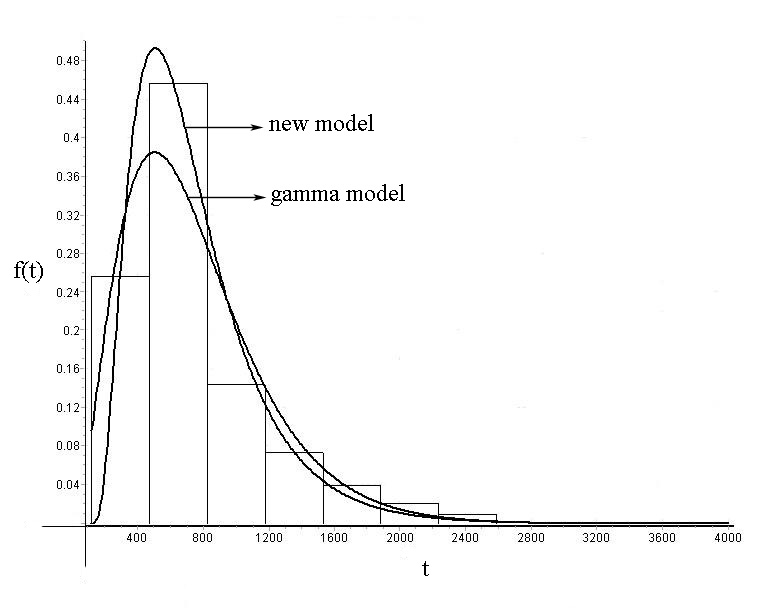}}\\
\caption{\small The graph of histogram embedded with the probability models.\label{ppplot5}}
\end{center}
\end{figure}

Our dataset consists of 1522 observations collected from  website\\ http://rredc.nrel.gov/solar/spectra/am0/.
Here, mathematical softwares MAPLE and MATLAB are used for the data analysis. In 2000, the American Society for Testing and Materials developed an
AM 0 reference spectrum (ASTM E-490) for use by the aerospace
community.  That ASTM E490 Air Mass Zero solar spectral irradiance
is based on data from satellites, space shuttle missions,
high-altitude aircraft, rocket soundings, ground-based solar
telescopes, and modeled spectral irradiance. The
model considered here is the density function given in (\ref{eq:4.3}). In many situations
gamma model is used to model the spectral density. The following
Figure \ref{ppplot5} is the histogram of the data embedded with gamma and
our new probability models. We haven't specified any parameters here to plot the
function. The same program generated the two different graphs as shown below.
We calculated Kolmogorov-Smirnov test statistic for the two different probability models. For
gamma density, the value of the statistic is obtained as 0.11139 and for our new probability model, the value is
0.10808. From the table, the value obtained is 0.410. We can see that the two
different probability models are consistent with the data. But the distance measure
of the statistic of our new probability model is less than the other probability
model and hence our model is better fit to the data than the other one.

\vskip.3cm \noindent
{\bf \large Acknowledgement }\\
\\
Author acknowledges gratefully the encouragements given by Professor A. M. Mathai, Department of Mathematics and Statistics, McGill University, Montreal,
Canada H3A 2K6.

\end{document}